# Effects of surface functional groups on the formation of nanoparticle-protein corona

R. Podila[1], R. Chen[2], P. C. Ke [2, 3] and A. M. Rao [2,3]

1. Department of Pharmacology and Toxicology, Brody School of Medicine, East Carolina University, Greenville, NC, USA 27834.
2. Department of Physics and Astronomy, Clemson University, Clemson, SC, USA 29634.
3. Center for Optical Materials Science and Technology, Clemson University, Clemson, SC, USA 29634.

ABSTRACT: Herein, we examined the dependence of protein adsorption on the nanoparticle surface in the presence of functional groups. Our UV-visible spectrophotometry, transmission electron microscopy, infrared spectroscopy and dynamic light scattering measurements evidently suggested that the functional groups play an important role in the formation of nanoparticle-protein corona. We found that uncoated and surfactant-free silver nanoparticles derived from a laser ablation process promoted a maximum protein (bovine serum albumin) coating due to increased changes in entropy. On the other hand, BSA displayed a lower affinity for electrostatically stabilized nanoparticles due to the constrained entropy changes.

Nanotechnology has been extensively used in the recent past for developing novel therapeutic methodologies.[1-7] Many nanostructured materials including carbon nanotubes (CNTs), noble metal nanoparticles (Au and Ag NPs) and lipid-based nanocomplexes (liposomes) have been proposed as effective drug delivery agents. Particularly, Ag NPs are being used widely for drug delivery, wound dressings, textile fabrics and wood flooring due to their anti-microbial properties.[4-5] Although considerable progress has been achieved with regards to NP synthesis, characterization, and drug loading, there are several challenges which continue to hinder the practical realization of nanostructured drug-delivery.[3,8,9] For example, the removal of NPs by macrophages poses a major hurdle for their *in vivo* application. It is therefore crucial to understand the mechanisms by which NPs interact with biomolecules, and consequently alleviate some of the existing technical challenges.[2,3]

It has been previously established that the reticuloendothelial system integrates NPs due to the adsorption of certain serum/complement proteins called 'opsonins'.[3,8,9] Upon exposure to serum or plasma, NPs rapidly adsorb protein molecules leading to the formation of a protein "corona".[2,10-12] The adsorption of proteins on NPs occurs mainly due to: i) an increase in the collective entropy of the proteins on an NP surface and ii) nonspecific interactions between the NP surface and the proteins.[13-15] In other words, the formation of protein corona has been found to depend largely on the physicochemical characteristics of the NPs, in addition to that of the proteins themselves.[10-12] Therefore, it is pertinent to understand the role of NP surface functionality in the formation of protein corona for exerting a control over NP-protein interactions. Understandably, optical spectroscopy is one of the key techniques for investigating the formation of protein corona since the shape and size of noble metal NPs can be probed from their surface plasmon resonance (SPR) while Fourier transform infrared (FTIR) spectroscopy can be employed to study the secondary structure of proteins. Here, we combined the

SPR and FTIR spectroscopy techniques to study the effect of surface functional groups on the formation of protein corona. We investigated the interactions between bovine serum albumin (BSA) and Ag NPs coated with different functional groups. To probe the effect of surface stabilizing agents or functional groups, we studied three independent systems: i) novel laser ablation (LA) produced Ag NPs directly dispersed in de-ionized water without the need for special stabilizing agents, ii) citric acid stabilized Ag NPs and iii) polyvinylpyrrolidone (PVP) stabilized Ag NPs. Interestingly, we found that the stabilizing agents such as citric acid and PVP indeed played an important role in the corona formation. Our FTIR and SPR spectra suggested that the proteins could rearrange readily to a more stable position on an uncoated Ag NP surface, resulting in an increased change in the conformational entropy ($\Delta S$). Such an increase in $\Delta S$ led to a significant decrease in the Gibbs free energy change ($\Delta G$), and promoted strong interactions between uncoated Ag NPs and the proteins.

Silver nanoparticles (citrate and PVP-coated Ag NPs) of 20 nm in diameter were purchased from NanoComposix and used in our experiments. Uncoated Ag NPs were synthesized using a laser ablation method in de-ionized (DI) water. One of the major advantages of laser ablation produced NPs is their extremely high colloidal stability due to their inherent surface charge, leading to stabilizing agent free Ag NPs. A silver metal foil target immersed in 10 ml of DI water was ablated with the 1064 nm beam (650 mJ) of a Nd:YAG laser operated at 10 Hz with a Q-switch delay of 175 μs. The as obtained suspension of Ag NPs was filtered using a 0.1 μm polytetrafluoroethylene membrane to obtain a final suspension. The surface chemistry of laser-ablated metal NPs is reported to be dominated by negatively charged functionalities like OH$^-$ and O$^-$ from DI water, which account for electrostatic stabilization of as produced particles.[16] It is important to note that, unlike citrate- and PVP-coated Ag NPs, the laser-ablated NPs were uncoated and did not possess any additional functional groups (apart from the ones resulted during the synthesis) or surfactants on their surfaces. BSA proteins were purchased from Sigma-Aldrich. The NPs and NP-BSA complexes were characterized using a Hitachi H-7600 transmission electron microscope. The optical properties were studied using a Perkin-Elmer Lambda 900 UV-Visible spectrophotometer and a Bruker IFS/v66 FTIR spectrometer.

As shown in Fig. 1a, the TEM images confirmed that the uncoated Ag NPs produced *via* laser ablation (LA Ag NPs) were spherical in shape and exhibited a unimodal diameter distribution with a mean value ~24 nm. Similarly, the average diameter of citrate- and PVP-coated Ag NPs was found to be ~20 nm, in agreement with the data provided by the vendor. For protein binding experiments, desired amounts of BSA (0-15 nM) were added to the Ag NP suspension (~10 μM) and thoroughly mixed *via* stirring to achieve a good dispersion. Next, the dispersions were incubated at 37 $^o$C for 1 hr. Subsequently, their UV-visible spectra were measured in a 10 mm quartz cuvette.

All Ag NPs exhibited the characteristic SPR peak at ~400 nm in the absence of BSA, which is consistent with previous studies,[17] (see Fig. 2). As it will be discussed later, the SPR peak for uncoated LA Ag NPs was found to be comparatively higher (~5 nm) to citric acid or PVP coated Ag NPs due to the differences in their immediate dielectric environment. This SPR peak is further redshifted with increasing exposure to BSA, and can be understood as follows. In metal NPs, the characteristic peak of SPR is dependent on the size, shape, and dielectric function ($\varepsilon$) of the NPs and its surrounding medium. In our case, the interaction of Ag NPs with BSA molecules resulted in a change of $\varepsilon$ which resulted in a redshift of the extinction coefficient maximum, and the observed redshift in the SPR peaks.[17] Interestingly, Fig. 2d showed that the amount of redshift is dependent on the nature of surface groups present on the Ag NPs. It is noteworthy that the amount of redshift did not seem to depend upon the initial concentration of BSA since: i) adsorption equilibrates at the microscopic time scales and ii) the amount of protein coating on NPs is much less than the total BSA content suggesting that all the BSA concentrations result in similar final coating. Clearly, the redshift was more pronounced for LA Ag NP-BSA relative to the citrate- and PVP-coated Ag NPs-BSA complex despite their similar changes in $\varepsilon$. Such variations in the redshifts can be explained in terms of the coating thickness of BSA. Previously, Ferenandez et al. observed that an increasing redshift occurred for Au NPs with their increasing thicknesses of silica coating.[18] Thus, in our case, the differences in the redshifts may be attributed to the different amounts of proteins adsorbed on the NP surfaces. This reasoning was further confirmed by our TEM studies. As shown in Figs. 1c-e, the LA Ag NPs exhibited more protein coating compared to citrate- and PVP-coated Ag NPs. The citrate- and PVP-coated Ag NPs aggregated upon exposure to proteins, indicating that their electrostatic stabilization groups were displaced from their surfaces. In agreement with the TEM results, the dynamic light scattering results indicated that the LA Ag NPs adsorbed the highest amount of BSA compared to the citrate-and PVP-coated Ag NPs (see Fig. 1b).

The change in enthalpy ($\Delta H$) for protein-NP interaction is usually small and negative. Thus, the Gibbs free energy decreases mainly as a result of the increasing $\Delta S$ ($\Delta G = \Delta H - T\Delta S$). As a protein molecule interacts with a NP surface, the energy of the system is minimized by the relaxation of protein secondary structures *via* the maximization of the protein entropy. The extent of protein conformational change is strongly dependent on the protein−NP surface interactions. Electrostatic interactions between the protein and NP surface could strongly impede the relaxation of the protein structure. To verify the mechanisms behind the variations in redshifts, we performed FTIR spectroscopy (Fig. 3). As shown in Fig. 3, the FTIR spectrum of BSA consists of several characteristic peaks. The so-called amide I region (∼1700−1600 cm$^{-1}$) arising from a C=O stretching vibration has been widely used for the identification of protein secondary structure.[19-21] The differences in secondary structures (α-helices, β-sheet, β-turn, and unordered) of the protein result in different vibrational frequencies for the C=O frequency. Thus, the overall shape and maxima of the amide I band can be effectively used to obtain the entropy of the protein upon its adsorption onto the NP surface. The peaks with maxima at ∼1690 cm$^{-1}$ are known to arise from β-sheets or β-turn structures, ∼1650 cm$^{-1}$ from α-helices, and 1630-1635 cm$^{-1}$ from extended or random coils.

To obtain the FTIR spectra of BSA coronas on the NPs, we centrifuged the incubated samples and washed the obtained pellets with DI water to remove any unadsorbed proteins. Also, we performed control experiments on BSA to ensure that the centrifugation/washing/drying steps did not induce any major changes in the FTIR spectra. The FTIR spectrum of each Ag NP-BSA composite was similar for all the BSA concentrations as in the case of the SPR spectra The FTIR spectrum for BSA exhibited all the three peaks for β-sheet or β-turn structures, α-helices, and extended or random coils (cf. Fig. 3). Upon adsorption of proteins on the Ag NPs, the peak ~1690 cm$^{-1}$ disappeared for all Ag NP types, suggesting that the protein entropy increased primarily *via* the relaxation of their β-sheets. The α-helices and random chains of the proteins could still be observed in the final corona, however. The ratios of the areas under random chains to α-helices ($I_{1630}/I_{1650}$) provided a quantitative measure for the net change in the protein entropy. As shown in Fig 4b, this ratio is < 1 for native BSA since it is more ordered, *i.e.,* more percentage of α-helices. Upon interacting with the NPs, the ratios evidently increased to more than 1, suggesting that all NP types induced an increase in the protein conformational entropy, or decrease the protein α-helix content. Interestingly, we found that the change in conformational entropy was the highest for the uncoated LA Ag NPs, consistent with the previously observed protein adsorption on bulk surfaces.[15] The BSA proteins could rearrange more readily to more stable positions onto an uncoated Ag NP surface leading to an increase in *ΔS* since long-range electrostatic interactions were much less pronounced in this case compared to the citrate- and PVP-coated Ag NPs. The hydrophobic interactions between uncoated Ag NPs and the proteins are also expected to play a stronger role in promoting their binding as compared to the corresponding interaction in the other two types of NPs. The overall effect of both greater entropic gain and stronger hydrophobic interactions favored more extensive coating of the proteins on LA Ag NPs, leading to the most significant redshift in the plasmon peak and likely the most stable NP-protein corona.

In summary, our UV-visible, TEM, and FTIR measurements evidently suggested that the increase in protein conformation entropy governed the extent of coating of BSA on Ag NPs. The net increase in $\Delta S$ was correlated with the nature of functional groups on the NP surfaces. Our FTIR measurements confirmed that Ag NP-BSA corona primarily relaxed the β-sheets, α-helices and increased the random coils of the proteins. Our SPR spectra, hydrodynamic size and the FTIR measurements indicated that the plasmon redshifts of Ag NPs correlated with the increased protein coating on the NPs. This study facilitates our understanding of the role of NP surface functionalization in protein corona formation, and has implications for nanomedicine, biotechnology, and nanotoxicology.

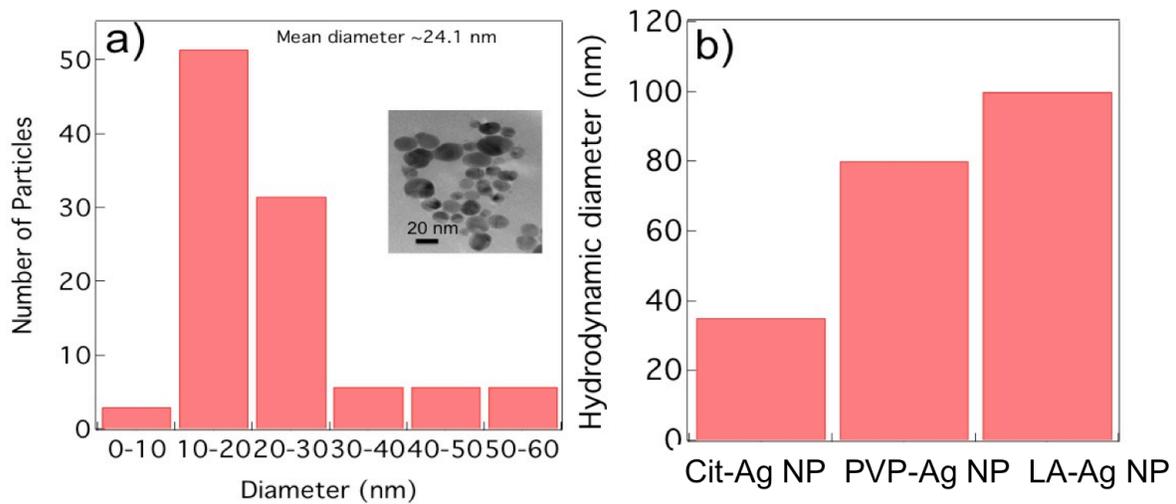
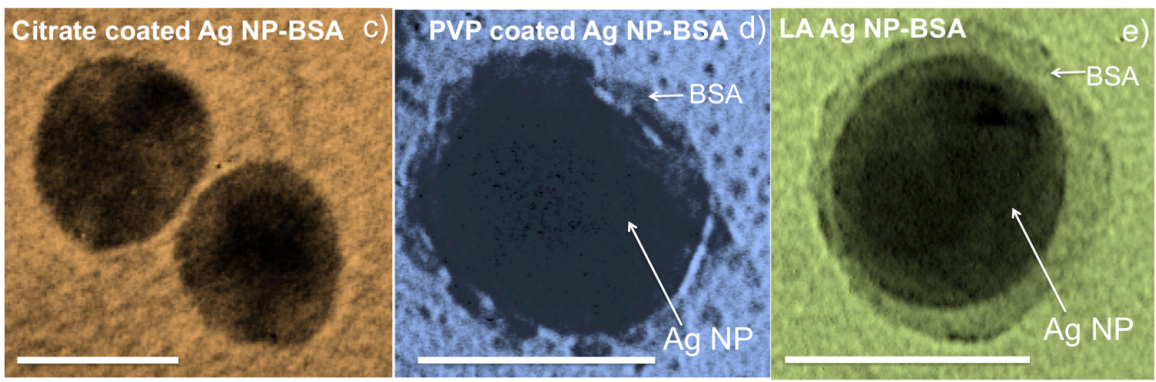

Figure 1: a) The diameter distribution for laser ablated Ag NPs exhibits a unimodal distribution with an average ~24 nm. The inset shows a representative TEM image for LA-Ag NPs b) The hydrodynamic diameter for BSA coated Ag NPs is different for different surface coatings. TEM images for BSA coated citrate coated Ag NPs (c) and PVP coated Ag NPs (d) and LA Ag NPs (e) show that the coating is maxium for LA Ag NPs in agreement with the hydrodynamic diameter. The scale bar for all the panels is 20 nm.

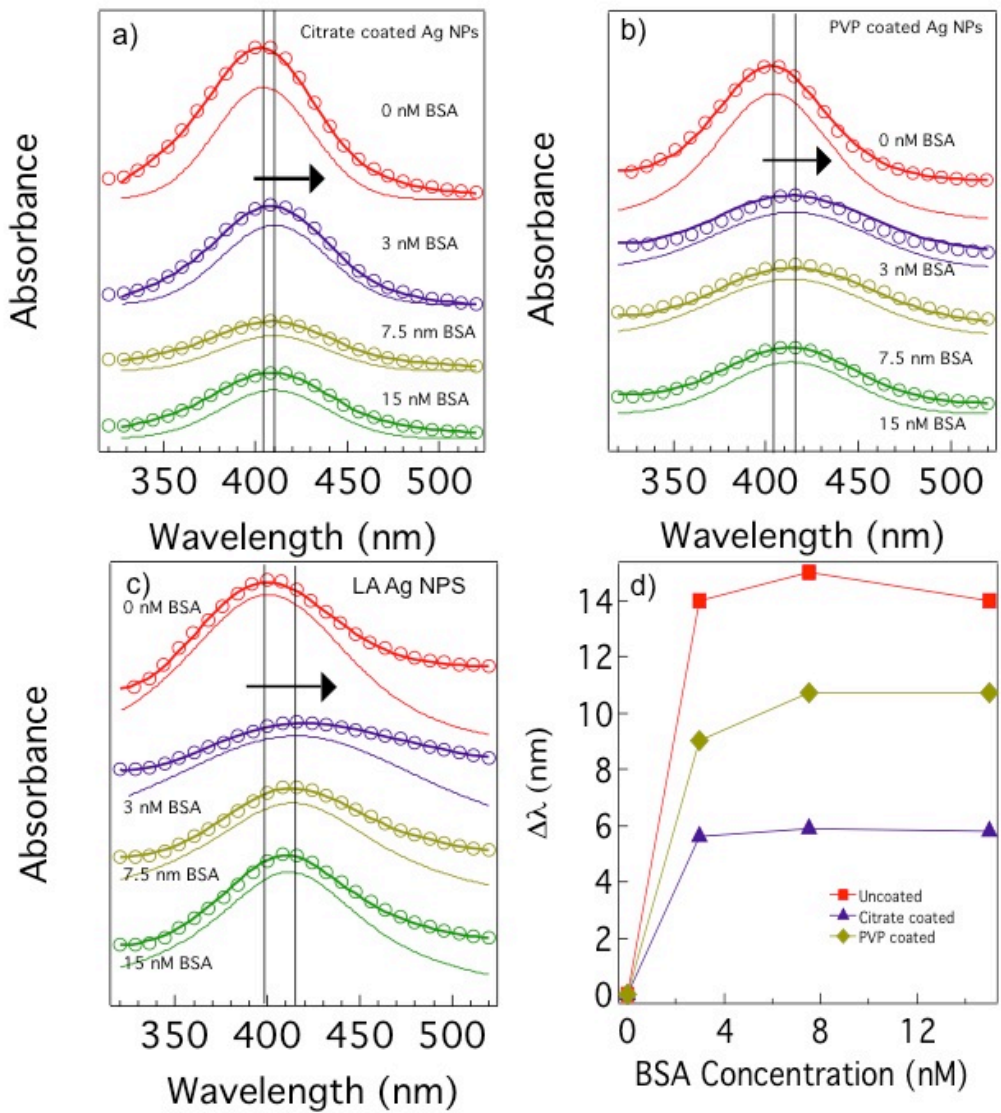

Figure 2: Surface plasmon spectra for citrate coated (a), PVP coated (b) and LA or uncoated (c) Ag NPs as a function of increasing BSA.(d) The redshift in the surface plasmon peak of Ag NPs as a function of BSA concentration.

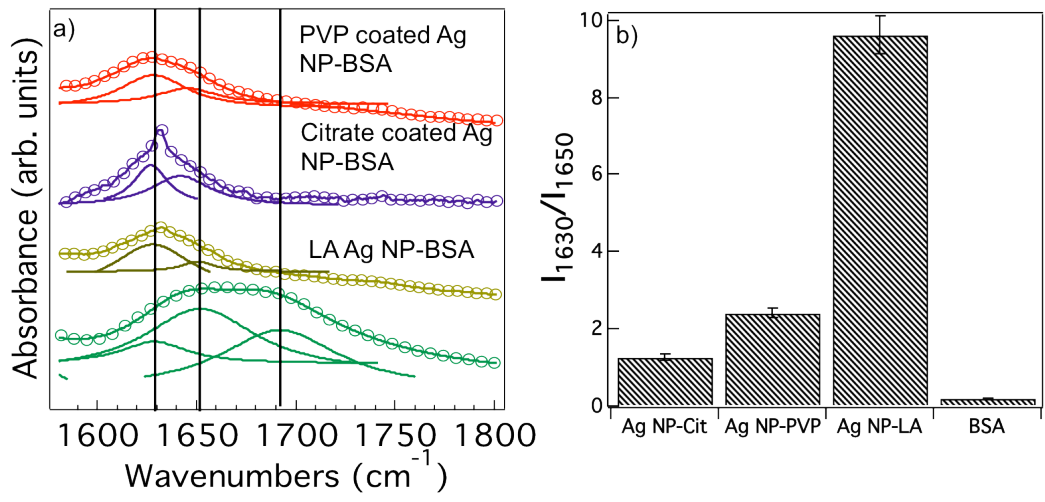

Figure 3: a) FTIR spectrum for native BSA and BSA coated LA Ag NPs, citrate-and PVP-coated Ag NPs. b) The ratio of the areas under random chains to α-helices ($I_{1630}/I_{1650}$) provides a quantitative measure for net change in the conformational entropy.